# Compositionally-warped additive mixed modeling for a wide variety of non-Gaussian spatial data

Daisuke Murakami[1,2*], Mami Kajita[1], Seiji Kajita[1], Tomoko Matsui[3]

[1] Singular Perturbations Co. Ltd., 1-5-6 Risona Kudan Building, Kudanshita, Chiyoda, Tokyo, 102-0074, Japan

[2] Department of Statistical Data Science, Institute of Statistical Mathematics, 10-3 Midori-cho, Tachikawa, Tokyo, 190-8562, Japan

[3] Department of Statistical Modeling, Institute of Statistical Mathematics, 10-3 Midori-cho, Tachikawa, Tokyo, 190-8562, Japan

* Correspondence: dmuraka@ism.ac.jp

Abstract: As with the advancement of geographical information systems, non-Gaussian spatial data is getting larger and more diverse. Considering this background, this study develops a general framework for fast and flexible non-Gaussian regression, especially for spatial/spatiotemporal modeling. The developed model, termed the compositionally-warped additive mixed model (CAMM), combines an additive mixed model (AMM) and the compositionally-warped Gaussian process to model a wide variety of non-Gaussian continuous data including spatial and other effects. Specific advantages of the proposed CAMM requires no explicit assumption of data distribution unlike existing AMMs, and fast estimation through a restricted likelihood maximization balancing the modeling accuracy and complexity. Monte Carlo experiments show the estimation accuracy and computational efficiency of CAMM for modeling non-Gaussian data including fat-tailed and/or skewed distributions. Finally, the proposed approach is applied to crime data to examine the empirical performance of the regression analysis and prediction. The proposed approach is implemented in an R package spmoran. See details on how to implement CAMM, see https://github.com/dmuraka/spmoran.

# 1. Introduction

A wide variety of spatial and spatiotemporal data is now becoming available. In addition to conventional spatial data collected and published in a top-down manner (e.g., census statistics), an increasing number of sensing data (e.g., remotely sensed images, smart sensor data) and other spatial data assembled, estimated, and disseminated by private companies and volunteers are available in the era of open data (Volunteered Geographic information; see Haklay, 2013). Such databases include Google Earth Engine (https://earthengine.google.com/), WorldPop (https://www.worldpop.org/), and Worldometer (https://www.worldometers.info/). This rapid growth of spatiotemporal open data drives people to use regression modeling to reveal hidden factors behind social issues such as crime occurrence (e.g., Kajita and Kajita, 2020) the spread of COVID-19 (e.g., Sannigrahi et al., 2020).

Despite the fact that spatial data are becoming increasingly diverse, Gaussian spatial regression models, which have been commonly used (e.g., Wikle and Cressie, 2011), are applicable only to data obeying a Gaussian distribution. Because regression problems are commonplace in spatial fields, an alternative approach for a wide variety of non-Gaussian spatial data is required.

According to Yan et al. (2020), representative modeling approaches for non-Gaussian spatial data are classified with (a) variable transformation and (b) generalized linear modeling. The former converts explained variables, which have a non-Gaussian distribution, to Gaussian variables through a transformation function. Logarithmic transformation (Dowd, 1982), Box-Cox transformation (Kitanidis and Shen, 1996), Tukey g-and-h transformation (Xu and Genton, 2017), and other transformation functions have been used in geostatistics (see Cressie and Wikle, 2011). In machine learning literature, the warped Gaussian process (WGP; Snelson et al., 2003) is a general framework including aforementioned transformation approaches. It has been extended to Bayesian WGP (Lázaro-Gredilla, 2012) and the compositionally-warped GP (CWGP; Rios and Tober, 2019), which we will focus on later.

The generalized linear model (b) is widely used for spatial modeling as well. This model accommodates spatial random effects and has been developed and extended under Bayesian framework (Gotway and Stroup, 1997; Diggle et al., 1998). While Bayesian approaches can be slow because of the simulation step, fast Bayesian inference including the integrated nested Laplace approximation (Rue et al., 2009) and Vecchia-Laplace approximation (Zilber and Katzfuss, 2019) has been developed for fast non-Gaussian spatial regression modeling. The generalized additive (mixed) model (e.g., Wood, 2017) is another popular approach that is applicable for non-Gaussian spatial data modeling (see, Umlauf, 2012). Computationally efficient estimation algorithms for the generalized additive model, including the fast restricted maximum likelihood (REML; Wood, 2011) and the separation of anisotropic penalties algorithm (Rodríguez-Álvarez et al., 2015) have been developed and implemented in a wide variety of software packages (see Mai and Zhang, 2018 for review).[1]

A critical limitation of the above-mentioned approaches include the need to assume data distribution a priori.[2] Because the true distribution behind data is usually unknown in empirical studies, distribution assumption can lead model misspecification. Exceptionally, CWGP, which is categorized in (a), estimates data distribution without explicitly assuming any distribution a priori (see Section 2.2). In short, CWGP is a distribution-free approach. Such an approach is potentially useful for modeling a wide variety of spatial and spatiotemporal data, although the original CWGP cannot consider spatial and/or temporal effects.

This study proposes an approach termed compositionally-warped additive mixed modeling (CAMM) as a unified regression approach for a wide variety of Gaussian and non-Gaussian continuous data. This approach is developed by combining CWGP and additive mixed models (AMM) that take roles to estimate data distribution without explicit prior information and quantify the spatial and other smooth and/or group effects depending on covariates, respectively. The remainder of the sections is organized as follows. Section 2 and 3 introduce AMM and CWGP respectively. Then, Section 4 develops CAMM by combining them. Section 5 performs Monte Carlo experiments to examine coefficients-estimation accuracy and computational efficiency of the developed approach. Section 6 employs CAMM to crime analysis in Tokyo, Japan. Finally, Section 7 concludes our discussion.

## 2. Additive mixed model (AMM)

AMM is a regression model accommodating spatial, temporal, and many other effects. The linear AMM is defined as follows:

$$\mathbf{y} = \mathbf{X}\boldsymbol{\beta} + \sum_{k=1}^{K} \mathbf{b}_k + \boldsymbol{\varepsilon}, \ldots \ldots \boldsymbol{\varepsilon} \sim N(\mathbf{0}, \sigma^2 \mathbf{I}), \qquad (1)$$

where $\mathbf{y} = [y_1, \ldots, y_N]'$ is a $N \times 1$ vector of explained variables, $\mathbf{X}$ is a $N \times J$ matrix of $J$ covariables assuming fixed effects, $\boldsymbol{\beta}$ is a $J \times 1$ vector of fixed coefficients. " ′ " represents the matrix transpose. $\sigma^2$ is the variance parameter, $\mathbf{0}$ is a zero vector, and $\mathbf{I}$ is an identity matrix. $K$ is the number of random effects.

$\mathbf{b}_k$ is a vector of random effects depending on the $k$-th variate $\mathbf{z}_k$, which is either a matrix or a vector of $k$-th covariate; for example, $\mathbf{b}_k$ describes a temporal process if $\mathbf{z}_k$ is defined by a $N \times$

---

[1] There are non-Gaussian spatial modeling approaches other than (a) and (b), including copula-based spatial modeling (e.g., Gräler, 2014), the Gaussian mixture approach (e.g., Fonseca and Steel, 2011), and approaches assuming non-Gaussian spatial processes (e.g., Zhang and El-Shaarawi, 2010). See Yan et al. (2020) for a literature review.

[2] In (a), a transformation function implicitly assumes data distribution. For example, the log-linear model assumes explained variables obey a log-normal distribution.

1 vector of temporal coordinates whereas spatial process if it is defined by a $N \times 2$ matrix of spatial coordinates. $\mathbf{b}_k$ is specified as follows:

$$\mathbf{b}_k = \mathbf{E}_k \boldsymbol{\gamma}_k, \qquad \boldsymbol{\gamma}_k \sim N(\mathbf{0}_k, \mathbf{V}_k(\boldsymbol{\theta}_k)), \tag{2}$$

where $\mathbf{E}_k$ is a $N \times L$ matrix of $L$ basis functions generated from $\mathbf{z}_k$. Orthogonal, radial, and other functions are available to generate basis functions from $\mathbf{z}_k$. $\boldsymbol{\gamma}_k$ is a vector of random coefficients depending on the variance-covariance matrix $\mathbf{V}_k(\boldsymbol{\theta}_k)$ and a set of variance parameters $\boldsymbol{\theta}_k$.

$\mathbf{b}_k$ can model group effects, non-linear effects, time-varying effects, and spatially varying effects by specifying $\mathbf{E}_k$ (see Umlauf et al., 2012). For example, if $\mathbf{E}_k$ is defined by spatial basis functions generated from spatial coordinates, $\mathbf{b}_k$ yields a spatial process. Moran eigenvectors (Griffith, 2003), which are extracted from a doubly-centered spatial proximity matrix, are spatial basis functions based on the Moran coefficient (MC; Moran, 1950). MC takes a positive value in the presence of positive spatial dependence while the value is negative in the presence of negative spatial dependence. The Moran eigenvectors corresponding to positive eigenvalue explain positively dependent map patterns (i.e., MC > 0) whereas those corresponding to negative eigenvalue explain negatively dependent patterns (i.e., MC < 0).

Because positive spatial dependence is dominant in most regional data (Griffith, 2003), the eigenvectors corresponding to positive eigenvalue are usually used to estimate underlying positively dependent spatial processes. Murakami and Griffith (2015) used the following Moran eigenvector-based specification for modeling positively dependent spatial process:

$$\mathbf{b}_k^{(s)} = \mathbf{E}_k^{(s)} \boldsymbol{\gamma}_k, \qquad \boldsymbol{\gamma}_k \sim N(\mathbf{0}_k, \tau_k^2 \boldsymbol{\Lambda}^{\alpha_k}), \tag{2}$$

where $\boldsymbol{\theta}_k \in \{\tau_k^2, \alpha_k\}$. $\mathbf{E}_k^{(s)}$ is a $N \times L$ matrix of the $L$ Moran eigenvectors corresponding to positive eigenvalue and $\boldsymbol{\Lambda}$ is a $L \times L$ diagonal matrix of the positive eigenvalues. $\tau_k^2$ estimates the variance of the spatial process while $\alpha_k$ estimates the Moran coefficient value or scale of the process (see Murakami and Griffith, 2020a).

In spatial statistics, it is a major concern to analyze spatially varying relationships between explained and explanatory variables (e.g., Brunsdon et al., 1998). Spatially varying coefficients (SVC) or other varying coefficients can be estimated using AMM by slightly modifying Eq. (1) as follows:

$$\mathbf{y} = \mathbf{X}\boldsymbol{\beta} + \sum_{k=1}^{K} \mathbf{x}_k \circ \mathbf{b}_k + \boldsymbol{\varepsilon}, \ldots \boldsymbol{\varepsilon} \sim N(\mathbf{0}, \sigma^2 \mathbf{I}), \tag{3}$$

where $\mathbf{x}_k$ is the k-th column of $\mathbf{X}$, $\circ$ is an element-wise (Hadamard) product operator, and $\mathbf{b}_k$ is the $k$-th varying coefficients vector. Later, $\mathbf{b}_k$ is defined by $\mathbf{b}_k^{(s)}$ to model SVCs.

The AMMs (Eqs. 1 and 3) have the following expression:

$$\mathbf{y} = \mathbf{X}\boldsymbol{\beta} + \widetilde{\mathbf{E}}\widetilde{\boldsymbol{\gamma}} + \boldsymbol{\varepsilon}, \quad \ldots \widetilde{\boldsymbol{\gamma}} \sim N(\widetilde{\mathbf{0}}, \widetilde{\mathbf{V}}(\boldsymbol{\Theta})), \quad .\boldsymbol{\varepsilon} \sim N(\mathbf{0}, \sigma^2 \mathbf{I}), \tag{4}$$

where $\widetilde{\mathbf{E}} = [\mathbf{E}_1, \ldots, \mathbf{E}_K]$ for Eq.(1) and $\widetilde{\mathbf{E}} = [\mathbf{x}_k \circ \mathbf{E}_1, \ldots, \mathbf{x}_k \circ \mathbf{E}_K]$ for Eq.(3). $\boldsymbol{\Theta} \in \{\boldsymbol{\theta}_1, \ldots, \boldsymbol{\theta}_K\}$, $\widetilde{\boldsymbol{\gamma}} =$

$[\boldsymbol{\gamma}_1, \ldots, \boldsymbol{\gamma}_K]'$, $\widetilde{\mathbf{0}} = [\mathbf{0}_1, \ldots, \mathbf{0}_K]'$, and $\widetilde{\mathbf{V}}(\boldsymbol{\Theta})$ is a block diagonal matrix whose $k$-th block is $\mathbf{V}_k(\boldsymbol{\theta}_k)$.

The likelihood function is given as follows (see Bates and DebRoy, 2004):

$$L^{AMM}(\boldsymbol{\beta}, \boldsymbol{\Theta}, \sigma^2) = \int \frac{1}{\sqrt{(2\pi\sigma^2)^{(N+\tilde{L})}|\widetilde{\mathbf{V}}(\boldsymbol{\Theta})|}} \exp\left(-\frac{\left\|\mathbf{y} - \mathbf{X}\boldsymbol{\beta} - \widetilde{\mathbf{E}}\widetilde{\boldsymbol{\gamma}}\right\| + \widetilde{\boldsymbol{\gamma}}\widetilde{\mathbf{V}}(\boldsymbol{\Theta})^{-1}\widetilde{\boldsymbol{\gamma}}}{2\sigma^2}\right) d\widetilde{\boldsymbol{\gamma}}, \quad (5)$$

where $\tilde{L} = \sum_{k=1}^{K} L_k$, and the restricted likelihood function as

$$L_R^{AMM}(\boldsymbol{\Theta}, \sigma^2) = \int L^{AMM}(\boldsymbol{\beta}, \boldsymbol{\Theta}, \sigma^2) d\boldsymbol{\beta}. \quad (6)$$

We prefer the restricted maximum likelihood (REML) estimation maximizing Eq. (6) because of its high estimation accuracy and stability shown by Reiss and Ogden (2009) and Wood (2011).

## 3. Compositionally-warped Gaussian process (CWGP)

Snelson et al. (2003) developed the warped GP (WGP) transforming non-Gaussian explained variables $\mathbf{y}$ to Gaussian variables through a transformation (or warping) function $\varphi()$. For independent samples, the WGP is defined as

$$\boldsymbol{\varphi}_{\boldsymbol{\omega}}(\mathbf{y}) \sim N(\boldsymbol{\mu}, \sigma^2 \mathbf{I}), \quad (7)$$

where $\boldsymbol{\varphi}_{\boldsymbol{\omega}}(\mathbf{y}) = [\varphi_{\boldsymbol{\omega}}(y_1), \ldots, \varphi_{\boldsymbol{\omega}}(y_N)]'$, $\boldsymbol{\omega}$ denotes the parameters characterizing the transformation, and $\boldsymbol{\mu}$ is a mean vector. Logarithmic, Box-Cox, and other transformations are available for the $\varphi_{\boldsymbol{\omega}}(\cdot)$ function. Note that the log-normal kriging and trans-Gaussian kriging (see Cressie, 1993), assuming these transformations are particular types of WGP with a spatially dependent covariance matrix.

A drawback of WGP is that the $\varphi_{\boldsymbol{\omega}}(\cdot)$ function must be given beforehand even though the true data distribution (or transformation function) is unknown. To break the bottleneck, Rios and Tober (2019) proposed the CWGP specifying the transformation function by concatenating $D$ sub-transformation functions. The CWGP for independent samples is defined as

$$\begin{aligned}\boldsymbol{\varphi}_{\boldsymbol{\omega}}(\mathbf{y}) &\sim N(\boldsymbol{\mu}, \sigma^2 \mathbf{I}), \\ \boldsymbol{\varphi}_{\boldsymbol{\omega}}(\mathbf{y}) &= \boldsymbol{\varphi}_{\boldsymbol{\omega}_D}\left(\boldsymbol{\varphi}_{\boldsymbol{\omega}_{D-1}}(\cdots \boldsymbol{\varphi}_{\boldsymbol{\omega}_1}(\mathbf{y}))\right),\end{aligned} \quad (8)$$

where $\boldsymbol{\omega} \in \{\boldsymbol{\omega}_1, \ldots, \boldsymbol{\omega}_D\}$ and $\boldsymbol{\varphi}_{\boldsymbol{\omega}_d}(\mathbf{y}) = [\varphi_{\boldsymbol{\omega}_d}(y_1), \ldots, \varphi_{\boldsymbol{\omega}_d}(y_N)]'$ represents the $d$-th transformation. Interestingly, they showed that CWGP approximates a wide variety of non-Gaussian distributions without explicitly assuming any non-Gaussian distribution a priori if each transformation is defined by the following function:

$$\varphi_{\boldsymbol{\omega}_d}(y_i) = \omega_{d,1} + \omega_{d,2}\sinh(\omega_{d,3}\operatorname{arcsinh}(y_i) - \omega_{d,4}), \quad (9)$$

where $\boldsymbol{\omega}_d \in \{\omega_{d,1}, \omega_{d,2}, \omega_{d,3}, \omega_{d,4}\}$. Based on Rios and Tober (2019), we call Eq.(9) as SAL transformation (Sinh-Arcsinh and Affine where the "L" comes from linear).

The likelihood for the CWGP model Eq.(8) yields

$$L(\mathbf{\mu}, \mathbf{\omega}, \sigma^2) = \frac{1}{\sqrt{(2\pi\sigma^2)^N}} \exp\left(-\frac{||\mathbf{\varphi_\omega}(\mathbf{y}) - \mathbf{\mu}||}{2\sigma^2}\right) \prod_{i=1}^{N} \frac{\partial \varphi_\omega(y_i)}{\partial y_i} \tag{10}$$

where

$$\frac{\partial \varphi_\omega(y_i)}{\partial y_i} = \frac{\partial \varphi_{\omega_D}(\varphi_{\omega_{D-1}}(\cdots \varphi_{\omega_1}(y_i)))}{\partial y_i} \cdots \frac{\partial \varphi_{\omega_2}(\varphi_{\omega_1}(y_i))}{\partial y_i} \frac{\partial \varphi_{\omega_1}(y_i)}{\partial y_i}. \tag{11}$$

Given Eq. (9), the *d*-th gradient in Eq. (11) yields

$$\frac{\partial \varphi_{\omega_d}(y_i)}{\partial y_i} = \frac{\omega_{d,2}\omega_{d,3}\cosh(\omega_{d,3}\operatorname{asinh}(y_i) - \omega_{d,4})}{\sqrt{1 + y_i^2}} \tag{12}$$

Based on Eq. (10), the log-likelihood becomes the sum of the Gaussian log-likelihood, whose computational complexity is $O(N)$ for independent samples and $O(N^3)$ in general, and the Jacobian term $\log\left(\prod_{i=1}^{N} \frac{\partial \varphi_\omega(y_i)}{\partial y_i}\right)$ whose complexity equals $O(N)$. In other words, CWGP has the same computational complexity as the classical GP. Thus, CWGP is a computationally efficient and flexible approach for non-Gaussian data modeling.

As a preliminary analysis, we test if the CWGP model in Eq. (8) accurately transforms non-Gaussian variables to Gaussian variables. Here we assume the following distributions of **y**: beta distribution (shape parameters: {2, 2}), the skew t-distribution (scale parameter: 6, slant parameter: 3), and the mixture of three Gaussian distributions with means {-2, 5, 10}, and variances {1, 1, 1}. In each case, **y** is defined by a vector of 1,000 random samples. For the mixture of Gaussians, 500 samples are generated from the mean 10 distribution whereas the other 250 and 250 samples are generated from mean -2 and 5 distributions respectively. For each **y**, CWGP is fitted by maximizing the likelihood. We assume $D \in \{1,2,4\}$.

Figure 1 summarizes the fitting results. This figure shows that CWGP accurately transforms the non-Gaussian distributions to Gaussian distributions even with a small number of transformations *D*. The accuracy CWGP is confirmed.

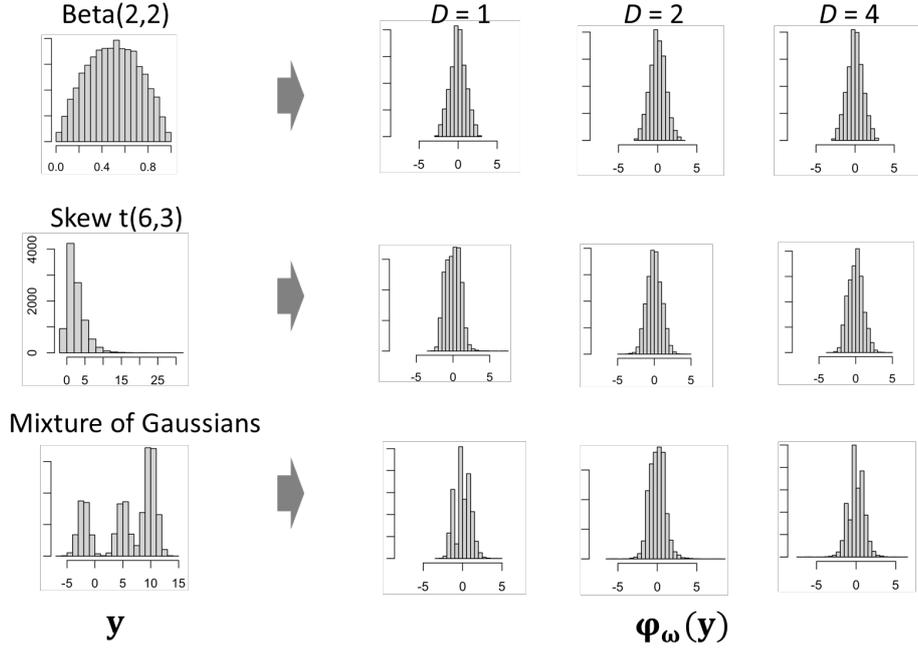

Figure 1: Result of CWGP fitting for beta distribution, skew t distribution, and Gaussian mixtures. Left three panels represent histograms of the simulated data, and the right nine panels show the histograms after the transformation. $D$ is the number of transformations.

## 4. Development of the compositionally-warped additive mixed modeling (CAMM) approach

### 4.1. Model

This section combines CWGP and AMM, and develops the CAMM for non-Gaussian data modeling without explicitly assuming any distribution.

The CAMM is defined as

$$\boldsymbol{\varphi_\omega}(\mathbf{y}) = \mathbf{X}\boldsymbol{\beta} + \sum_{k=1}^{K} \mathbf{x}_k \circ \mathbf{b}_k + \boldsymbol{\varepsilon}, \dots \dots \boldsymbol{\varepsilon} \sim N(\mathbf{0}, \sigma^2 \mathbf{I}), \tag{13}$$

with

$$\boldsymbol{\varphi_\omega}(\mathbf{y}) = \boldsymbol{\varphi}_{\omega_D}\big(\boldsymbol{\varphi}_{\omega_{D-1}}(\cdots \boldsymbol{\varphi}_{\omega_1}(\mathbf{y}))\big), \tag{14}$$

$$\mathbf{b}_k = \mathbf{E}_k \boldsymbol{\gamma}_k, \qquad \boldsymbol{\gamma}_k \sim N(\mathbf{0}_k, \mathbf{V}_k(\boldsymbol{\theta}_k)). \tag{15}$$

CAMM describes a wide variety of non-Gaussian explained variables through the transformation Eq. (14). While we consider Eq. (13) assuming the varying coefficients $\beta_k \mathbf{1} + \mathbf{b}_k$ on $\mathbf{x}_k$, CAMM can consider many other effects by specifying the random effects term $\mathbf{b}_k$.

### 4.2. Likelihood functions

As with other (C)WGPs, the likelihood function for our model is readily obtained by adding the Jacobian term $\prod_{i=1}^{N}\frac{\partial \varphi_\omega(y_i)}{\partial y_i}$ with the likelihood $L^{AMM}(\boldsymbol{\beta}, \boldsymbol{\Theta}, \sigma^2)$ of the linear AMM as follows:

$$L(\boldsymbol{\beta}, \boldsymbol{\Theta}, \boldsymbol{\omega}, \sigma^2)$$
$$= \int \frac{1}{\sqrt{(2\pi\sigma^2)^{(N+\tilde{L})}|\widetilde{\mathbf{V}}(\boldsymbol{\Theta})|}} \exp\left(-\frac{\left\|\boldsymbol{\varphi}_\omega(\mathbf{y}) - \mathbf{X}\boldsymbol{\beta} - \widetilde{\mathbf{E}}\tilde{\boldsymbol{\gamma}}\right\| + \tilde{\boldsymbol{\gamma}}\widetilde{\mathbf{V}}(\boldsymbol{\Theta})^{-1}\tilde{\boldsymbol{\gamma}}}{2\sigma^2}\right)\prod_{i=1}^{N}\frac{\partial\varphi_\omega(y_i)}{\partial y_i}d\tilde{\boldsymbol{\gamma}}. \quad (16)$$

Remember that $\widetilde{\mathbf{E}} = [\mathbf{x}_k \circ \mathbf{E}_1, \dots, \mathbf{x}_k \circ \mathbf{E}_K]$. Because $\prod_{i=1}^{N}\frac{\partial\varphi_\omega(y_i)}{\partial y_i}$ does not include $\tilde{\boldsymbol{\gamma}}$, the likelihood is further expanded as (see Eq. 5)

$$L(\boldsymbol{\beta}, \boldsymbol{\Theta}, \boldsymbol{\omega}, \sigma^2) = L^{AMM}(\boldsymbol{\beta}, \boldsymbol{\Theta}, \boldsymbol{\omega}, \sigma^2)\prod_{i=1}^{N}\frac{\partial\varphi_\omega(y_i)}{\partial y_i}, \quad (17)$$

where $L^{AMM}(\boldsymbol{\beta}, \boldsymbol{\Theta}, \boldsymbol{\omega}, \sigma^2)$ is identical to $L^{AMM}(\boldsymbol{\beta}, \boldsymbol{\Theta}, \sigma^2)$ in Eq. (5) except that $\mathbf{y}$ is replaced with $\boldsymbol{\varphi}_\omega(\mathbf{y})$:

$$L^{AMM}(\boldsymbol{\beta}, \boldsymbol{\Theta}, \boldsymbol{\omega}, \sigma^2)$$
$$= \int \frac{1}{\sqrt{(2\pi\sigma^2)^{(N+\tilde{L})}|\widetilde{\mathbf{V}}(\boldsymbol{\Theta})|}} \exp\left(-\frac{\left\|\boldsymbol{\varphi}_\omega(\mathbf{y}) - \mathbf{X}\boldsymbol{\beta} - \widetilde{\mathbf{E}}\tilde{\boldsymbol{\gamma}}\right\| + \tilde{\boldsymbol{\gamma}}\widetilde{\mathbf{V}}(\boldsymbol{\Theta})^{-1}\tilde{\boldsymbol{\gamma}}}{2\sigma^2}\right)d\tilde{\boldsymbol{\gamma}}. \quad (18)$$

On the other hand, the restricted likelihood function for the CAMM is given as

$$\begin{aligned}L_R(\boldsymbol{\Theta}, \boldsymbol{\omega}, \sigma^2) &= \int L^{AMM}(\boldsymbol{\beta}, \boldsymbol{\Theta}, \boldsymbol{\omega}, \sigma^2)\prod_{i=1}^{N}\frac{\partial\varphi_\omega(y_i)}{\partial y_i}d\boldsymbol{\beta}, \\ &= \int L^{AMM}(\boldsymbol{\beta}, \boldsymbol{\Theta}, \sigma^2)d\boldsymbol{\beta}\prod_{i=1}^{N}\frac{\partial\varphi_\omega(y_i)}{\partial y_i} = L_R^{AMM}(\boldsymbol{\Theta}, \boldsymbol{\omega}, \sigma^2)\prod_{i=1}^{N}\frac{\partial\varphi_\omega(y_i)}{\partial y_i},\end{aligned} \quad (19)$$

where $L_R^{AMM}(\boldsymbol{\Theta}, \boldsymbol{\omega}, \sigma^2)$ equals $L_R^{AMM}(\boldsymbol{\Theta}, \sigma^2)$ that has $\boldsymbol{\varphi}(\mathbf{y})$ instead of $\mathbf{y}$.

The restricted log-likelihood, which is maximized in the REML, becomes

$$\log L_R(\boldsymbol{\Theta}, \boldsymbol{\omega}, \sigma^2) = \log L_R^{AMM}(\boldsymbol{\Theta}, \boldsymbol{\omega}, \sigma^2) + \sum_{i=1}^{N}\log\frac{\partial\varphi_\omega(y_i)}{\partial y_i}. \quad (20)$$

The computational efficiency heavily depends on the computational complexity of $\log L_R^{AMM}(\boldsymbol{\Theta}, \boldsymbol{\omega}, \sigma^2)$ whereas the complexity of $\sum_{i=1}^{N}\log\frac{\partial\varphi_\omega(y_i)}{\partial y_i}$ is $O(N)$, which is quite small as same as the original CWGP.

### 4.3. Fast REML

For fast parameter estimation, we exploit the property that, given $\boldsymbol{\omega}$, $\log L_R^{AMM}(\boldsymbol{\Theta}, \boldsymbol{\omega}, \sigma^2)$ is identical to the restricted log-likelihood for the linear AMM (Eqs. 1 and 3). For the AMM, Murakami and Griffith (2019; 2020a) developed a fast REML relying on pre-conditioning. The pre-conditioning replaces the matrices $\{\mathbf{y}, \mathbf{X}, \widetilde{\mathbf{E}}\}$ whose size grows depending on the sample size *N*, with small inner products whose sizes do not grow even if *N* increases. Owing to this property, the fast REML is quite fast and memory efficient even for very large samples.

To make CAMM feasible for very large samples, we extend their algorithm to maximize the restricted log-likelihood). Specifically, we maximize it to estimate the parameters $\{\boldsymbol{\Theta}, \boldsymbol{\omega}, \sigma^2\}$ as follows:

(1) Initialize the parameter values $\{\widehat{\boldsymbol{\Theta}}, \widehat{\boldsymbol{\omega}}\}$.

(2) Evaluate the inner product matrices $\widetilde{\mathbf{M}} \in \{\mathbf{M}_{XX}, \mathbf{M}_{EX}, \mathbf{M}_{EE}, \mathbf{m}_{Xy}, \mathbf{m}_{Ey}, m_{yy}\}$ where $\mathbf{M}_{XX} = \mathbf{X}'\mathbf{X}$, $\mathbf{M}_{EX} = \widetilde{\mathbf{E}}'\mathbf{X}$, $\mathbf{M}_{EE} = \widetilde{\mathbf{E}}'\widetilde{\mathbf{E}}$, $\mathbf{m}_{Xy} = \mathbf{X}'\boldsymbol{\varphi}_{\widehat{\omega}}(\mathbf{y})$, $\mathbf{m}_{Ey} = \widetilde{\mathbf{E}}'\boldsymbol{\varphi}_{\widehat{\omega}}(\mathbf{y})$, $m_{yy} = \boldsymbol{\varphi}_{\widehat{\omega}}(\mathbf{y})'\boldsymbol{\varphi}_{\widehat{\omega}}(\mathbf{y})$.

(3) Update $\widehat{\boldsymbol{\Theta}}$ by numerically estimating $\widehat{\boldsymbol{\theta}}_k = \mathrm{argmax}_{\boldsymbol{\theta}_k} \log L_R(\boldsymbol{\theta}_k | \widehat{\boldsymbol{\Theta}}_{-k}, \widehat{\boldsymbol{\omega}}, \widehat{\sigma}^2, \widetilde{\mathbf{M}})$ sequentially for each *k*, where $\widehat{\boldsymbol{\Theta}}_{-k} = \{\widehat{\boldsymbol{\theta}}_1, \ldots, \widehat{\boldsymbol{\theta}}_{k-1}, \widehat{\boldsymbol{\theta}}_{k+1}, \ldots, \widehat{\boldsymbol{\theta}}_K\}$.

(4) Update $\widehat{\boldsymbol{\omega}}$ though maximization of $\log L_R(\boldsymbol{\omega}|\widehat{\boldsymbol{\Theta}}, \widehat{\sigma}^2, \widetilde{\mathbf{M}})$. It is done by iterating the following calculations:

   (i) Update $\widetilde{\mathbf{M}} \in \{\ldots, \mathbf{m}_{Xy}, \mathbf{m}_{Ey}, \mathbf{m}_{yy}\}$ where $\mathbf{m}_{xy} = \mathbf{X}'\boldsymbol{\varphi}_{\widehat{\omega}}(\mathbf{y})$, $\mathbf{m}_{Ey} = \widetilde{\mathbf{E}}'\boldsymbol{\varphi}_{\widehat{\omega}}(\mathbf{y})$, $m_{yy} = \boldsymbol{\varphi}_{\widehat{\omega}}(\mathbf{y})'\boldsymbol{\varphi}_{\widehat{\omega}}(\mathbf{y})$.

   (ii) Evaluate $\log L_R(\boldsymbol{\omega}|\widehat{\boldsymbol{\Theta}}, \widehat{\sigma}^2, \widetilde{\mathbf{M}})$.

(5) End if the log-restricted likelihood $\log L_R(\widehat{\boldsymbol{\Theta}}, \widehat{\boldsymbol{\omega}}, \widehat{\sigma}^2, \widetilde{\mathbf{M}})$ value converges. Otherwise, go back to step (3).

Owing to the pre-conditioning step (2) replacing large matrices with small inner product matrices, the heaviest calculation required in the numerical maximization steps (3) – (4) is the evaluation of $\widetilde{\mathbf{E}}'\boldsymbol{\varphi}_{\widehat{\omega}}(\mathbf{y})$. Fortunately, the computational complexity is *O*(*N*) that is reasonably small and feasible for large samples.

However, the steps (3) – (4) evaluating the restricted log-likelihood ($\log L_R^{AMM}(\boldsymbol{\Theta}, \boldsymbol{\omega}, \sigma^2) + \sum_{i=1}^{N} \log \frac{\partial \varphi_{\omega}(y_i)}{\partial y_i}$) can be slow if the number *K* of random effects is large (e.g., *K* = 10). To address the problem, we employ the fast REML of Murakami and Griffith (2019; 2020a) estimating the linear AMM computationally efficiently even when *N* and *K* are large. Because $\boldsymbol{\theta}_k$ is not in the Jacobian term $\sum_{i=1}^{N} \log \frac{\partial \varphi_{\omega}(y_i)}{\partial y_i}$, step (3) estimating $\boldsymbol{\theta}_k$ is achieved by maximizes $\log L_R^{AMM}(\boldsymbol{\Theta}, \boldsymbol{\omega}, \sigma^2)$ that is identical to the log-likelihood of the linear AMM; the fast REML is available. Although the Jacobian terms must be considered when estimating $\widehat{\boldsymbol{\omega}}$ in step (4), their fast algorithm is still applicable to evaluate $\log L_R^{AMM}(\boldsymbol{\Theta}, \boldsymbol{\omega}, \sigma^2)$ while the computational cost of $\sum_{i=1}^{N} \log \frac{\partial \varphi_{\omega}(y_i)}{\partial y_i}$ is marginal. All the likelihood evaluation steps are computationally efficient.

Later, the computational efficiency of the proposed estimation algorithm is examined through Monte Carlo experiments.

## 4.4. Practical issues

### 4.4.1. Specification of the transformation function

At times, explained variables are constrained by non-negativity or other upper/lower limits. Such constraints can be considered in the first transformation $\varphi_{\omega_1}(y_i)$. For example, the Box-Cox transformation, which is defined below, is useful to transform non-negative variables to unconstraint variables, which the SAL transformation implicitly assumes:

$$\varphi_{\omega_1 = \lambda}(y_i) = \begin{cases} \frac{y_i^\lambda - 1}{\lambda} & if\ \lambda \neq 0 \\ \log(y_i) & if\ \lambda = 0 \end{cases}, \tag{21}$$

where $\lambda$ is a parameter determining the non-linearity of the transformation. The Box-Cox transformation yields a linear transformation if $\lambda = 1$ while log-transformation if $\lambda = 0$. The log-transformation is available for the same purpose. The log-ratio of other transformations, which has been used for compositional data analysis (e.g., Egozcue et al., 2003), are available to convert variables with a box constraint to unconstrained variables. The first transformation must be specified carefully considering such value constraints in explained variables.

It is well known that the logarithmic and Box-Cox transformations are not available if explained variables have zero values. For instance, when $\mathbf{y} = [0, 0.001, 0.5, 1, 2]'$, the logged variables yield $\boldsymbol{\varphi}_\omega(\mathbf{y}) = [-\text{Inf}, -6.91, -0.69, 0.00, 0.69]$ (no SAL transformation assumed here). Because of the singular value, the residual sum of squares $||\boldsymbol{\varphi}_\omega(\mathbf{y}) - \mathbf{X}\boldsymbol{\beta} - \tilde{\mathbf{E}}\tilde{\boldsymbol{\gamma}}||$ in the likelihood takes an infinite value, which implies an extremely poor fit. In many real-world cases, the accuracy will be low even if the zero value is omitted from $\mathbf{y}$ because -6.91 is still extremely large relative to other logged values in the absolute sense. To avoid such difficulty, a small value has been added in practice. Such an addition can be regarded as another transformation before the log/Box-Cox transformation, and the small value can be estimated to maximize the likelihood under the CAMM framework (see Appendix 2). If the addition of a small value is added as the first transformation, log and Box-Cox transformations are available in our case even if the explained variables have zero values.

Although not assumed in Rios and Tober (2019), we introduce standardization before and after the SAL transformations to stabilize the estimation. The former is required to unify the initial condition and stabilize the REML estimation. The latter is required to stabilize $\boldsymbol{\varphi}_\omega(\mathbf{y})$; without the latter standardization, the intercept and regression coefficients tend to be extremely large, and the variance of $\boldsymbol{\varphi}_\omega(\mathbf{y})$ as well. Furthermore, in each SAL transformation, the non-negativity of $\omega_{d,2}$ and $\omega_{d,3}$ is also assumed to avoid the reversal of the large and small relationship in $\mathbf{y}$.

A remaining issue is the selection of $D$. Fortunately, Bayesian Information Criterion (BIC) or other criterion is available for the selection.

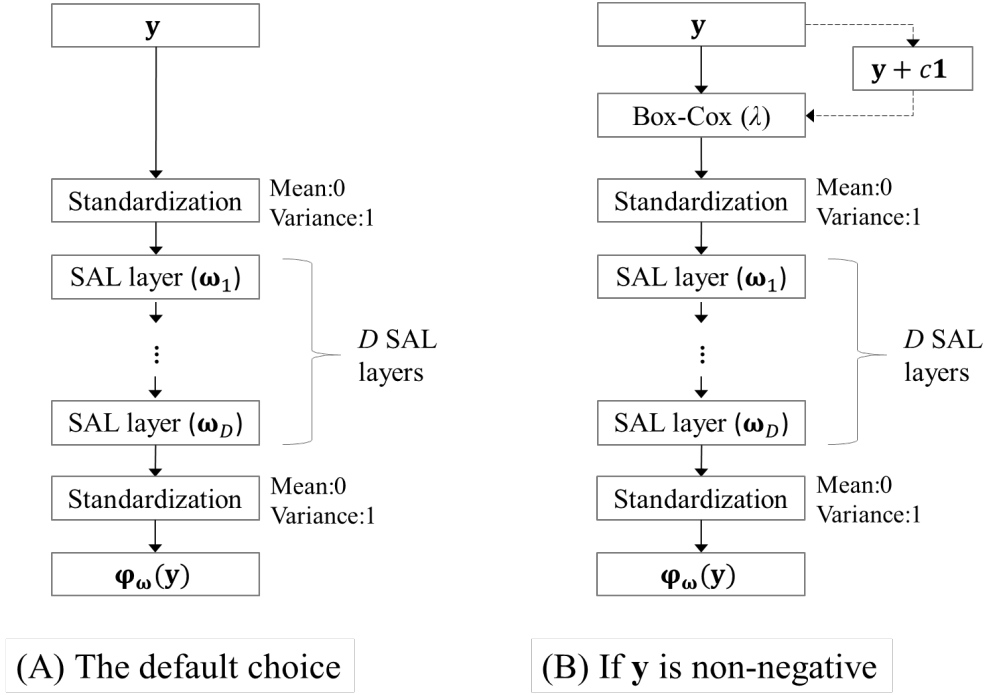

Figure 2: Transformation functions for CAMM. (A) is the suggested default function while (B) is recommended if **y** is non-negative. *c* is a small value used if **y** includes zero or very small values.

Finally, our assumed transformations are summarized in Figure 2. Remember that the developed approach assumes continuous explained variables. When modeling discrete variables like counts, they must be converted to density, for example, by dividing by area or population.

4.4.2. Interpretability of the estimation results

The estimated coefficients $\boldsymbol{\beta}$ and random effects $\mathbf{b}_1, \ldots, \mathbf{b}_K$ quantify the impact of $x_{i,k}$ on $\varphi_{\boldsymbol{\omega}}(y_i)$. While these estimates are directly interpretable just like in existing studies using log-transformed explained variables, the marginal effect $\partial y_i / \partial x_{i,k}$, which quantifies the influence on the raw explained variables $y_i$ is easier to interpret. The marginal effect is expanded as follows:

$$\frac{\partial y_i}{\partial x_{i,k}} = \frac{\partial y_i}{\partial \varphi_{\boldsymbol{\omega}}(y_i)} \frac{\partial \varphi_{\boldsymbol{\omega}}(y_i)}{\partial x_{i,k}}. \tag{22}$$

$\frac{\partial y_i}{\partial \varphi_{\boldsymbol{\omega}}(y_i)} = \left(\frac{\partial \varphi_{\boldsymbol{\omega}}(y_i)}{\partial y_i}\right)^{-1}$ where $\frac{\partial \varphi_{\boldsymbol{\omega}}(y_i)}{\partial y_i}$ was evaluated when evaluating the likelihood and $\frac{\partial \varphi_{\boldsymbol{\omega}}(y_i)}{\partial x_{i,k}} = \beta_k + b_{k,i}$ if Eq. (13) is considered. Thus, the marginal effect is easily evaluated. While $\frac{\partial y_i}{\partial x_{i,k}}$ changes depending on $y_i$, we recommend using the median, which is robust against outliers, as a summary statistic measuring the marginal effects from explained variables on explanatory variables.

# 5. Monte Carlo experiment

## 5.1. Setting

This section compares coefficient estimation accuracy and computational time through Monte Carlo experiments. Among coefficient specifications, we focus on spatially varying coefficients (SVCs) because of the following reasons: (i) SVC modeling is popular in spatial fields (see Fotheringham et al., 2003); (ii) SVC estimates tend to be unstable (e.g., Wheeler and Tiefelsdorf, 2005; Cho et al., 2009; Murakami et al., 2017); (iii) SVC estimation can be very slow depending on the estimation algorithm (Murakami and Griffith, 2019).

This study generates non-Gaussian explained variables from the following model:

$$y_i = \varphi_{g,h}(y_{0,i}) \quad y_{0,i} = \beta_{0,i} + x_{1,i}\beta_{1,i} + x_{2,i}\beta_{0,i} + \varepsilon_i, \quad \varepsilon_i \sim N(0, 2^2)$$

$$\beta_{0,i} = 1 + \sum_{j \neq i} w_{i,j} u_{0,j}, \quad u_{0,j} \sim N(0,1),$$

$$\beta_{1,i} = -2 + \sum_{j \neq i} w_{i,j} u_{1,j}, \quad u_{1,j} \sim N(0, 3^2), \tag{23}$$

$$\beta_{2,i} = 0.5 + \sum_{j \neq i} w_{i,j} u_{2,j}, \quad u_{2,j} \sim N(0,1),$$

where the explanatory variates $\{x_{1,i}, x_{2,i}\}$ are randomly generated from uniform distributions with the intervals [0, 1]. The SVCs $\{\beta_{0,i}, \beta_{1,i}, \beta_{2,i}\}$ are generated from the spatial moving average processes where $w_{i,j}$ is the $(i, j)$-th element of a matrix defined by row-standardizing a spatial proximity matrix whose $(i,j)$-th element equals $\exp(-d_{i,j}/0.5)$. Spatial coordinates used to evaluate the Euclidean distance $d_{i,j}$ are generated using two independent standard normal distributions.[3] $\beta_{1,i}$ is assumed to have a larger variation than the other two by assuming a larger variance than $\{\beta_{0,i}, \beta_{2,i}\}$.

$\varphi_{g,h}(\cdot)$ represents the Tukey g-and-h transformation (Tukey, 1977) that generates skewed and/or fat-tailed distributions, which is for $y_{0,i}$ is defined as

$$\varphi_{g,h}(y_{0,i}) = \frac{\exp(gy_{0,i} - 1)}{g} \exp\left(\frac{hy_{0,i}}{2}\right). \tag{24}$$

The parameters $g$ and $h$ control the skewness and kurtosis (or tail fatness) of the distribution of $\varphi_{g,h}(y_{0,i})$. A larger positive $g$ means stronger right skew while a larger negative $g$ means the opposite. A larger $h$ means a fatter tail distribution.

---

[3] In regional analysis, samples are typically concentrated in urban areas. This assumption implies urban areas with dense samples in the center.

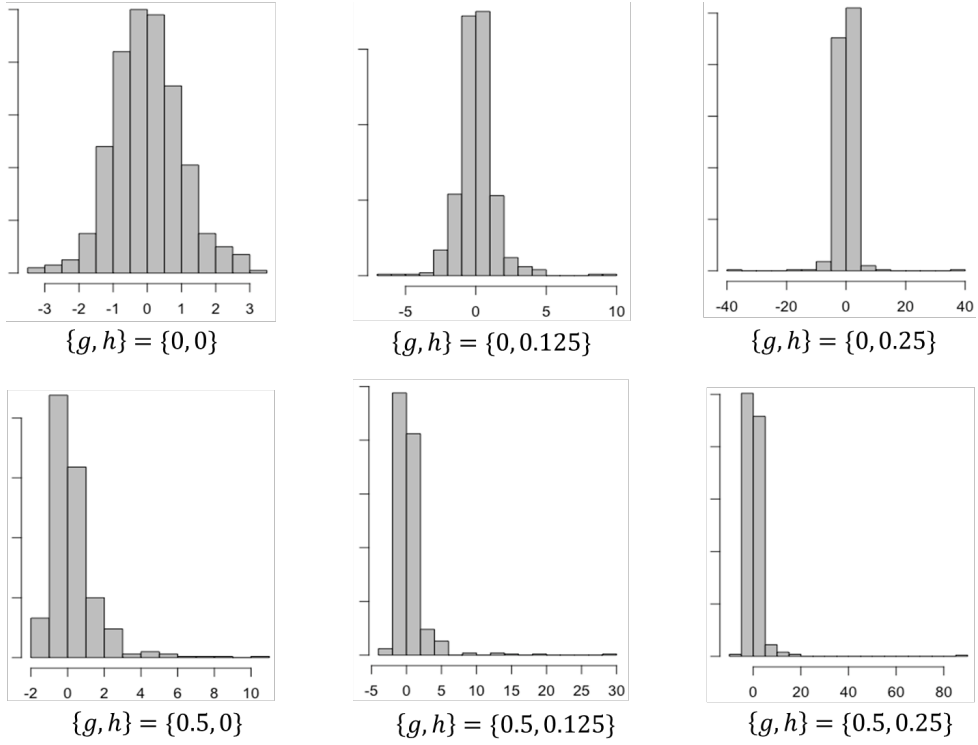

Figure 3: Histograms of the explained variables and the *g* and *h* values (*N* = 500)

The CAMMs with the number of SAL transformations being $D \in \{1, 2, 3, 4\}$ are compared with the linear regression (LM) and the linear additive mixed model (AMM), which replaces $\boldsymbol{\varphi}_\omega(\mathbf{y})$ with $\mathbf{y}$. For CAMMs and AMM, SVCs are assumed on $\{\beta_{0,i}, \beta_{1,i}, \beta_{2,i}\}$. These models were fitted 200 times while varying $g \in \{0.0, 0.5\}$, $h \in \{0, 0.125, 0.25\}$, and sample size $N \in \{50, 500, 5{,}000\}$. Figure 3 displays the histogram of the explained variables generated from the *g* and *h* values. This figure shows that our setting covers a wide range of skewed and/or fat tailed distributions. Note that we also performed simulations assuming $g = -0.5$ assuming left skew distributions. However, the result is quite similar to the results when $g = 0.5$; this is probably because distributions given in these two cases are symmetric. We do not report results when $g = -0.5$.

Throughout this section, we use R version 4.0.2 (https://cran.r-project.org/) installed in a Mac Pro (3.5 GHz, 6-Core Intel Xeon E5 processor with 64 GB memory).

5.2. Result: Estimation accuracy

The coefficient estimation accuracy is evaluated using the root mean squared error (RMSE), which is defined as follows:

$$RMSE(\beta_{k,i}) = \sqrt{\frac{1}{200N} \sum_{iter=1}^{200} \sum_{i=1}^{N} (\hat{\beta}_{k,i}^{(iter)} - \beta_{k,i})^2} \tag{25}$$

where *iter* represents the iteration number, and $\hat{\beta}_{i,p}^{(iter)}$ is the coefficient estimate in the *iter*-th iteration.

Figures 4 and 5 summarize the RMSE values of the estimated SVCs when $g = 0.0$ and $g = 0.5$, respectively. $\beta_{1,i}$, which has a larger spatial variation, is called strong SVC while $\beta_{2,i}$ is called weak SVC.

As expected, LM has poor estimation accuracy across cases. Figure 4 suggests that if the explained variables have Gaussian distribution ($g = h = 0.0$), both AMM and CAMMs accurately estimate the SVCs. However, the estimation error of AMM rapidly increases as skewness or kurtosis increases. Even worse, AMM estimation accuracy of the weak SVC ($\beta_{2,i}$) is worse than LM, possibly because of the weak identifiability. These results clearly suggest that AMM should not be used for non-Gaussian data.

In contrast, CAMMs have smaller RMSE values than AMM in all the cases including cases with $N = 50$. When $D \in \{2, 3, 4\}$, in each case, the RMSE values obtained from CAMM are small and similar to each other. These results verify that the proposed model accurately estimates coefficients and that the estimates are robust against the number of transformations if they are equal to or greater than 2.

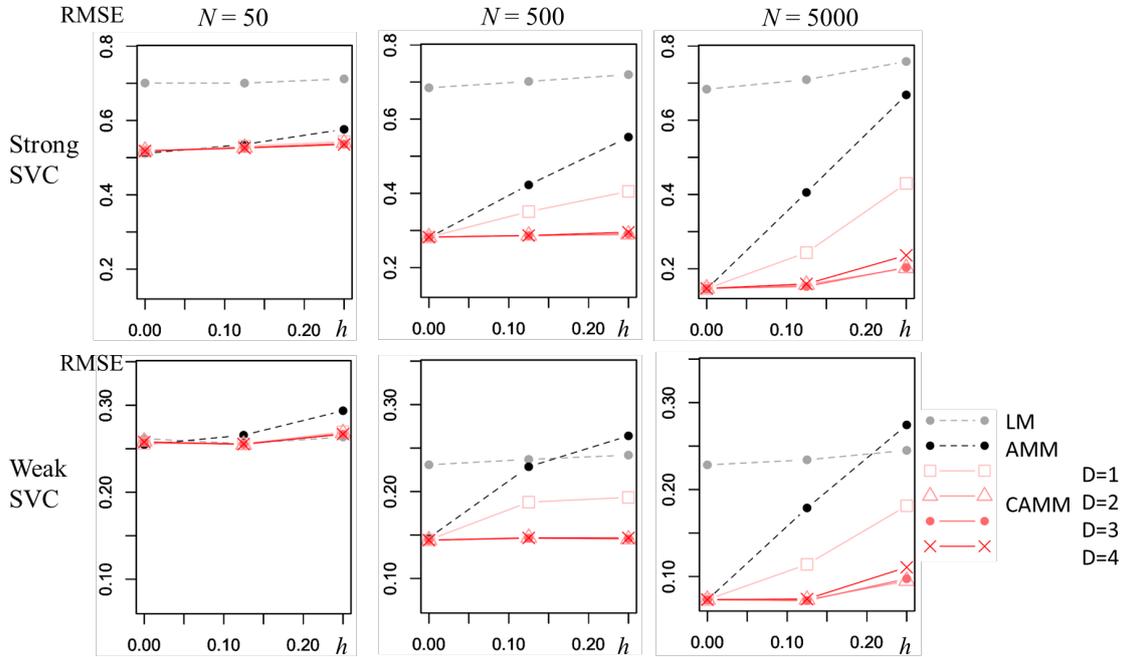

Figure 4: RMSE of the SVC estimates ($g = 0.0$)

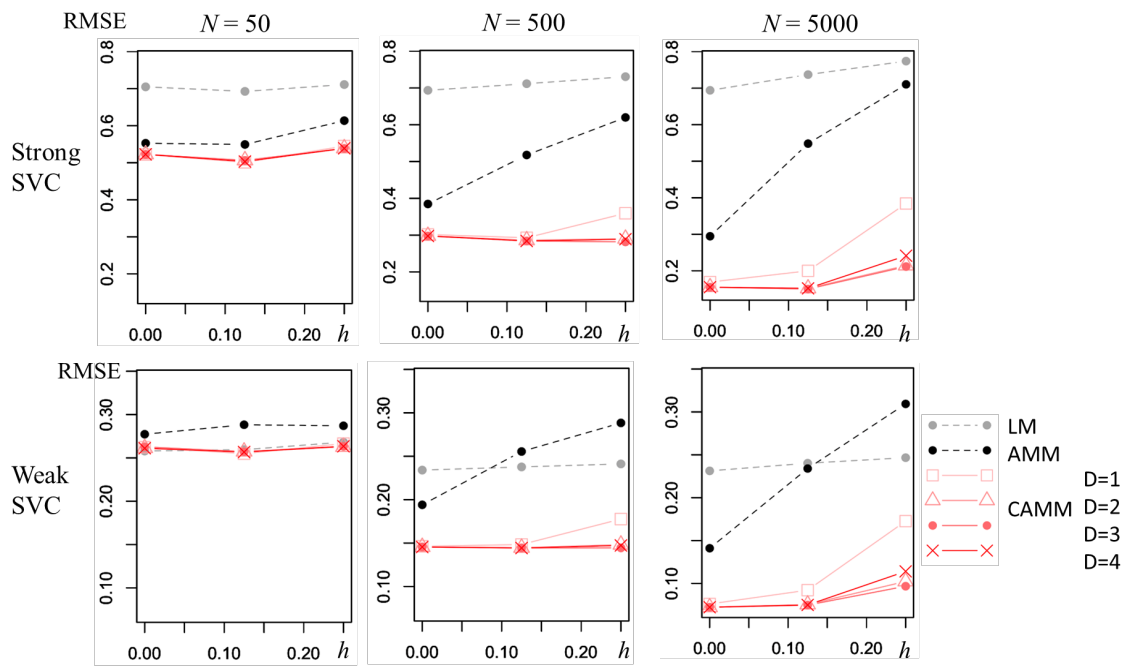

Figure 5: RMSE of the SVC estimates ($g = 0.5$)

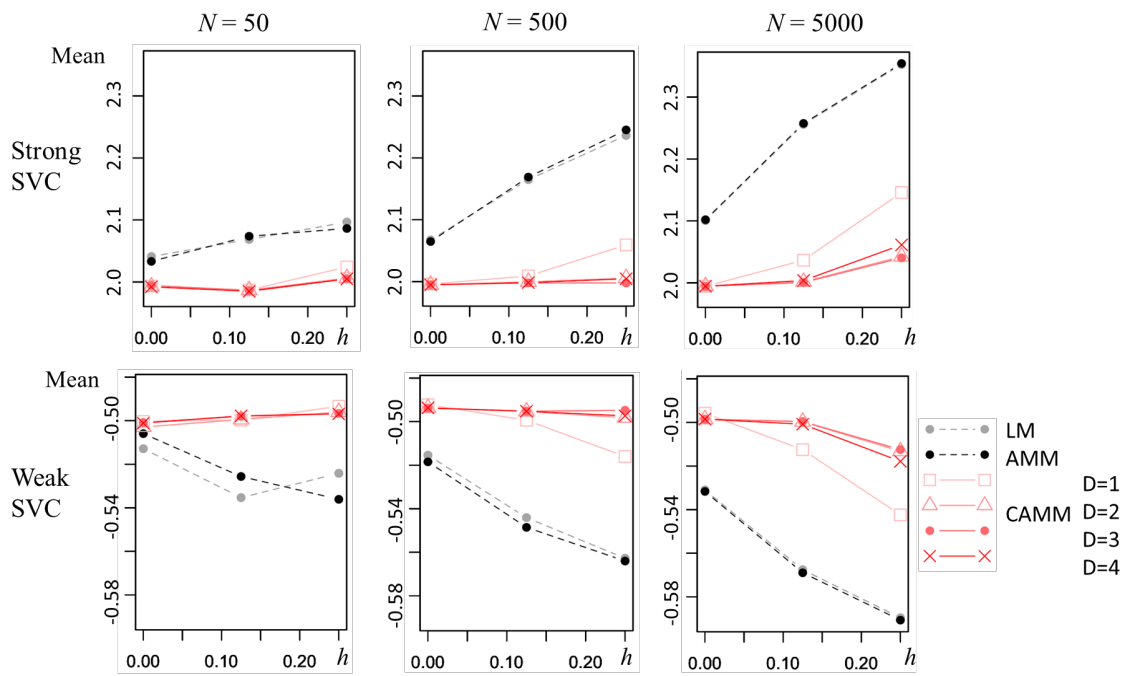

Figure 6: Mean of the SVC estimates ($g = 0.5$)

Finally, Figure 6 summarizes the mean of the SVC estimates, which is defined as

$$Mean(\beta_{k,i}) = \frac{1}{200N} \sum_{iter=1}^{200} \sum_{i=1}^{N} \hat{\beta}_{k,i}^{(iter)}. \tag{26}$$

The SVC estimates are unbiased if the mean is close to the true values that equal -2.0 for the strong SVC and 0.5 for the weak SVC. Figure 6 shows that the LM and AMM tend to underestimate the coefficients in absolute value. The CAMM estimates have much smaller biases for both the strong and weak SVCs. CAMM is found to be useful to estimate SVCs while avoiding underestimation of the effects from explanatory variables.

5.3. Result: Computation time

We then compare computation time assuming $g = 0.50$, $h = 0.25$, and $N \in \{1,000, 10,000, 20,000, 30,000, 40,000, 50,000\}$. In addition to LMM and AMM, CAMMs are compared with geographically weighted regression (GWR; Brunsdon et al., 1998), which is a popular SVC modeling approach. The GWmodel package (Lu et al., 2014a; version 2.1-4) is used for the estimation.[4]

Figure 7 summarizes the results. Of course, CAMMs are slower than the linear AMM since they additionally estimate $4D$ parameters numerically. Despite this, the computation time is fairly small across CAMMs even for large samples. For example, on average, the CAMM estimation with $D = 4$ took only 659 seconds for 50,000 samples. The computation time is considerably shorter than GWR despite CAMM numerically estimating $2K + 4D$ parameters while GWR numerically estimates only one bandwidth parameter. It is also found that, when $D \geq 2$, the computation time is similar even if $D$ is increased.

In short, Section 5 confirms estimation accuracy and computational efficiency of our proposed approach. In the next section, this approach is applied to a crime analysis in Japan.

---

[4] See Murakami et al., (2017) for the comparison of the SVC estimation accuracy between spatial AMM and GWR.

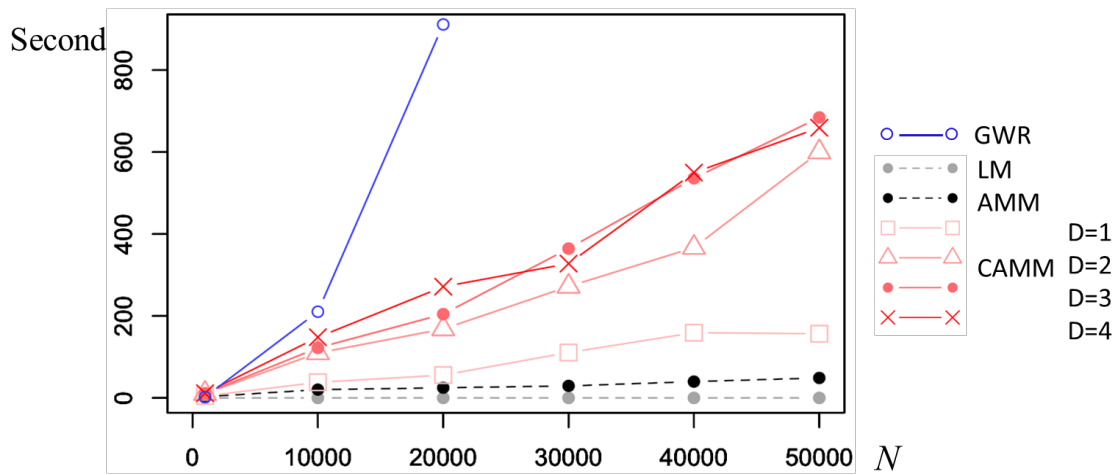

Figure 7: Computation time ($g = 0.5$; $h = 0.25$)

## 6. Application to crime modeling

6.1. Outline

This section compares the classical ANN with CAMM through crime modeling in Tokyo. We focus on shoplifting, which is a frequent non-burglary crime in this area, whose number of occurrences by minor municipal districts in Tokyo are provided in the Dai-Tokyo Bouhan network database (https://www.bouhan.metro.tokyo.lg.jp/), published by Promotion of Citizen Safety, Tokyo Metropolitan Government (https://www.tomin-anzen.metro.tokyo.lg.jp/english/). Roughly speaking, the area including and along the Yamanote line is the central area of Tokyo (see Figure 8).

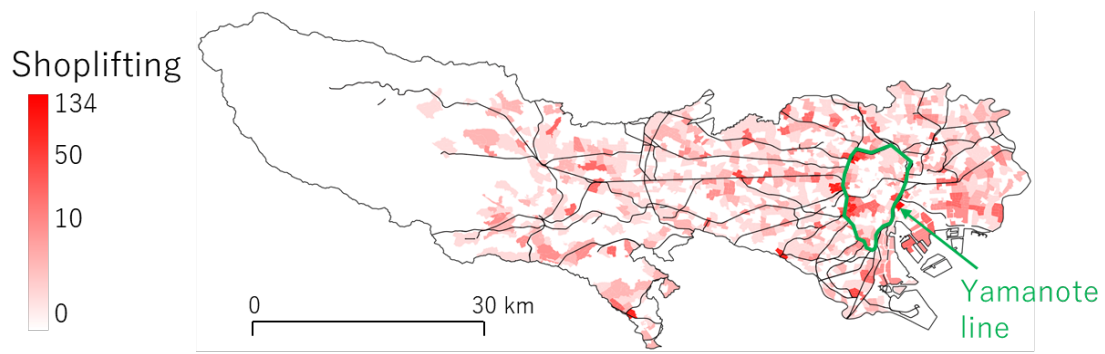

Figure 8: The number of shoplifting occurrences per area (the first quarter of 2017). The Yamanote line is a major railway surrounding the central area of Tokyo. The black lines denote the other railways except subways.

The explained variables are [the number of shoplifting occurrences] / [the number of retail stores] by minor municipal districts by quarter. The study period is between 2015 and 2018. The resulting sample size is 23,936. The explanatory variables are as listed in Table 1. Our model is specified as (see Murakami and Griffith, 2020b)

$$\varphi_{\omega}(y_i) = \sum_{k=1}^{K} x_{k,i}\ddot{\beta}_{k,i} + b_0(s_i) + b_0(t_i) + \varepsilon_k, \quad \varepsilon_k \sim N(0, \sigma^2). \quad (27)$$

where $\ddot{\beta}_{k,i}$ is the $k$-th coefficient for the $i$-th sample is specified as

$$\ddot{\beta}_{k,i} = \beta_k + b_k(s_i) + b_k(x_{k,i}). \quad (28)$$

$\beta_k$ is a constant, $b_k(s_i)$ is a spatially varying coefficient (SVC) where $s_i$ is the $i$-th sample site that is given by the geographic center coordinates of the districts, and $b_k(x_{k,i})$ is a non-spatially varying coefficient (NVC) smoothly varying depending on explanatory variable value. The SVC is given using Eq. (2) to model the positively dependent spatial process while the NVC is given by the natural spline functions generated from $x_{k,i}$. $b_k(s_i)$ and $b_k(x_{k,i})$ are accepted only if they improve the Bayesian information criterion (BIC); otherwise, they are replaced with zero values.

Table 1: Explanatory variables (source: Japan National Census 2015). See Murakami et al. (2020) for details about crime theory underlying the explanatory variables.

| Name | Description | Interpretation in terms of crime theory |
| --- | --- | --- |
| Repeat | Logged number of bicycle thefts per area in the previous quarter | They explain the tendency that crimes tend to repeat in similar regions due to regional heterogeneity (e.g., resident characteristics) and/or repetitive crimes by the same groups (see Farrell, 1995). |
| RepOther | Logged number of other non-burglary crimes in the previous quarter | |
| Popden | Nighttime population density [1000 people/km$^2$] | Based on the routine activity theory (Felson, 1994), (a) potential offenders and (b) suitable targets are triggers of crimes. Popden and Retail are used to quantify the density of (a) and (b) respectively. |
| Retail | Number of retails per area [Number/km$^2$] | |
| Fpopden | Ratio of foreigners among residents | Local environmental factors explaining race, economic deprivation, and education. |
| UnEmp | Unemployment ratio | |
| Univ | Ratio of residents who graduated university | |

In addition, we consider the time-varying intercept $b_0(t_i)$ by quarter $t_i$ to eliminate residual temporal dependence while the spatially varying intercept $b_0(s_i)$ defined by Eq.(2) to eliminate residual spatial dependence. They are important because an ignorance of residual spatial and/or temporal dependence can lead to incorrect inferences (see LeSage and Pace, 2009).

We apply AMM with log-transformed explained variables (AMM(log)), which is a popular specification in regional science, and CAMM with/without the Box-Cox transformation (i.e., A and B in Figure 2). The number of the SAL transformations is changed between 1 and 4 and compared the Bayesian information criterion (BIC) value to select the best CAMM specification. See Appendix. 3 for details about an implementation of CAMM using the R package "spmoran" (version 0.2.1 onward).

6.2. Transformation results

Figure 9 compares BIC across the AMM ($D = 0$) and CAMM ($D > 0$) specifications. This figure shows that the SAL transformations considerably improve the BIC value. The BIC value takes the minimum value when $D = 3$ without the Box-Cox transformation (BIC: -106,002) whereas $D = 2$ with the Box-Cox transformation (BIC: -109,747). It is particularly useful that the complexity of the transformation can be selected through a BIC minimization.

Later, we will compare coefficients estimated from the CAMM with $D = 2$ and the Box-Cox transformation, which we will call CAMM(BC+2SAL), with AMM(log) whose BIC equals -68,447.

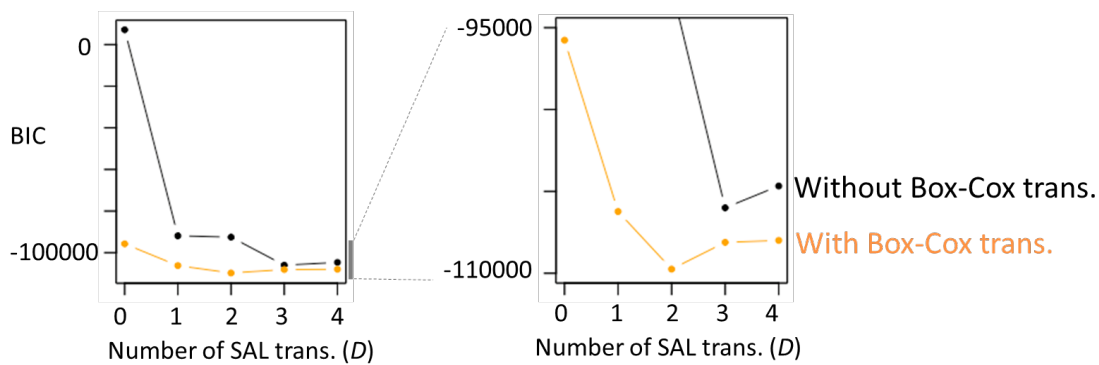

Figure 9: Comparison of the BIC values across AMM ($D = 0$) and CAMM ($D > 0$) specifications. The right panel is an enlarged view of the left panel.

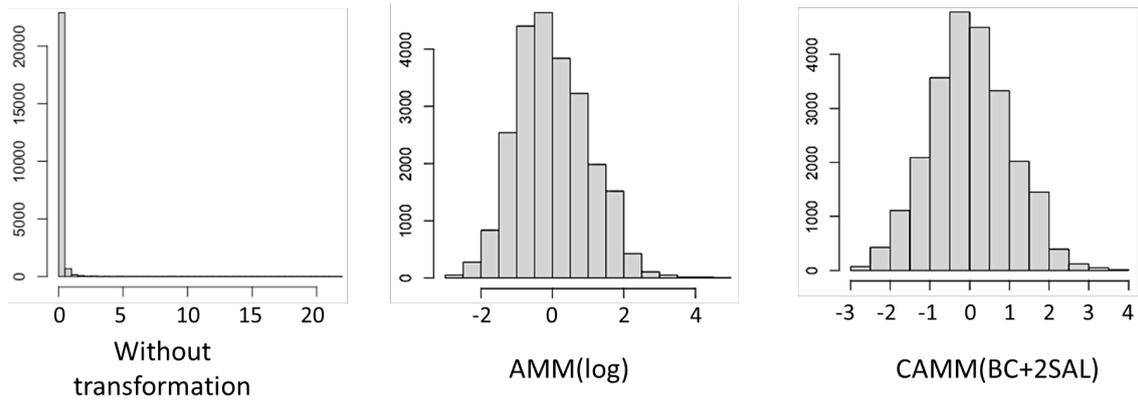

Figure 10: Histograms of the explained variables before and after the transformations

Figure 10 compares a histogram of the explained variables before and after the transformations. The histogram after the log-transformation, which AMM(log) assumes, is closer to the Gaussian distribution than the original histogram. Yet, the transformed distribution is left-skewed. The log- transformation might be insufficient in our case. In contrast, the histogram obtained through the estimation of CAMM(BC+2SAL) is much closer to the Gaussian distribution without explicit assumption of the distribution. The proposed transformation is conformed to improve the accuracy of the Gaussian approximation.

6.3. Parameter estimation result

This section compares the parameters estimated from AMM(log) with CAMM(BC+2SAL). Table 2 and 3 summarize the estimated coefficients and their statistical significances, respectively. As explained, the coefficient type is selected among constant, SVC, NVC, or SNVC (= SVC + NVC) through the BIC minimization. These tables suggest that estimation results are fairly similar. For both models, Fpopden, UnEmp, and Univ are statistically insignificant.

By contrast, Repeat is positively significant at the 1 % level in these models in most areas. Both models estimated that the coefficients on Repeat are SNVCs, which vary spatially and non-spatially depending on the Repeat value. Figure 11 plots the estimated coefficients on Repeat. The two models suggest that shoplifting is repeated in the central area of Tokyo, which is the major commercial area. CAMM(BC+2SAL) estimates a greater gap between the central and the other areas compared to the AMM(log) estimates. CAMM might be useful to accurately capture spatial heterogeneity in regression coefficients while coping with non-Gaussianity.

Table 2: Summary of Coefficient Estimates.

| | AMM(log) | | | | | | | |
|---|---|---|---|---|---|---|---|---|
| | Intercept | Repeat | RepOther | Popden | Retail | Fpopden | UnEmp | Univ |
| Selected coef. type | SVC | SNVC | | | Const | | | |
| Minimum | 1.938 | 0.660 | | | | | | |
| 1st quantile | 1.957 | 0.702 | | | | | | |
| Median | 1.970 | 0.712 | 0.092 | $-1.022 \times 10^{-3}$ | $8.932 \times 10^{-6}$ | -0.049 | 0.037 | 0.031 |
| 3rd quantile | 2.010 | 0.720 | | | | | | |
| Maximum | 2.082 | 0.762 | | | | | | |
| | CAMM(BC+2SAL) | | | | | | | |
| | Intercept | Repeat | RepOther | Popden | Retail | Fpopden | UnEmp | Univ |
| Selected coef. type | SVC | SNVC | SNVC | | Const | | | |
| Minimum | 1.872 | 0.534 | 0.032 | | | | | |
| 1st quantile | 1.894 | 0.670 | 0.088 | | | | | |
| Median | 1.906 | 0.687 | 0.099 | $-8.544 \times 10^{-4}$ | $1.358 \times 10^{-5}$ | -0.068 | 0.039 | 0.044 |
| 3rd quantile | 1.969 | 0.706 | 0.108 | | | | | |
| Maximum | 2.091 | 0.770 | 0.145 | | | | | |

**Table 3.** Proportion of statistical significance levels in each of the coefficients

| | AMM(log) | | | | | | | |
|---|---|---|---|---|---|---|---|---|
| Significance | Intercept | Repeat | RepOther | Popden | Retail | Fpopden | UnEmp | Univ |
| 10% level | 0.000 | 0.000 | 0.000 | 0.000 | | | | |
| 5% level | 0.000 | 0.000 | 0.000 | 0.000 | 0.000 | 0.000 | 0.000 | 0.000 |
| 1% level | 1.000 | 1.000 | 1.000 | 1.000 | | | | |
| | CAMM(BC+2SAL) | | | | | | | |
| Significance | Intercept | Repeat | RepOther | Popden | Retail | Fpopden | UnEmp | Univ |
| 10% level | 0.000 | 0.000 | 0.000 | 0.000 | 1.000 | | | |
| 5% level | 0.000 | 0.000 | 0.001 | 1.000 | 0.000 | 0.000 | 0.000 | 0.000 |
| 1% level | 1.000 | 1.000 | 0.999 | 0.000 | 0.000 | | | |

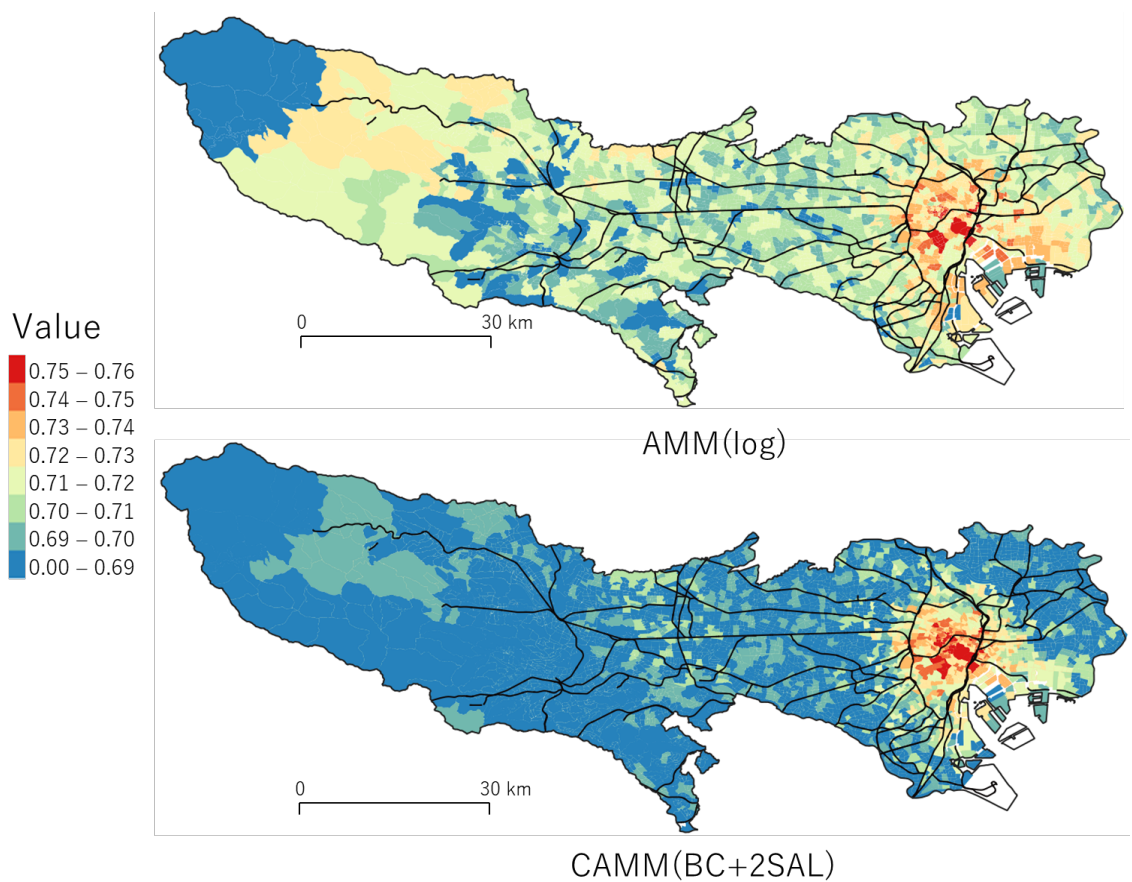

Figure 11: Estimated SNVC on Repeat. The black lines are railways not including subways.

RepOther is positively significant for both models. The estimated coefficient type is SNVC based on CAMM(BC+2SAL) while it is constant in AMM(log). Regarding their BIC values (see Fig. 9), the CAMM estimates are expected to be more reliable. The map of the CAMM coefficients suggests that the tendency of co-occurrence with other crimes is prominent in the cities of Hachioji and Haijima (Figure 12). They are local centers where people are concentrated. The large RepOther values in these cities might mean that shoplifting and other non-burglary crimes increase or decrease together depending on the characteristics of local people varying over time.

Based on CAMM(BC+2SAL), Popden is negatively significant at the 5 % level while Retail is positively significant at the 10 % level. These results show a higher risk of shoplifting in exclusive commercial areas, which are typically in the central area or nearby railway station areas. The result is intuitively reasonable.

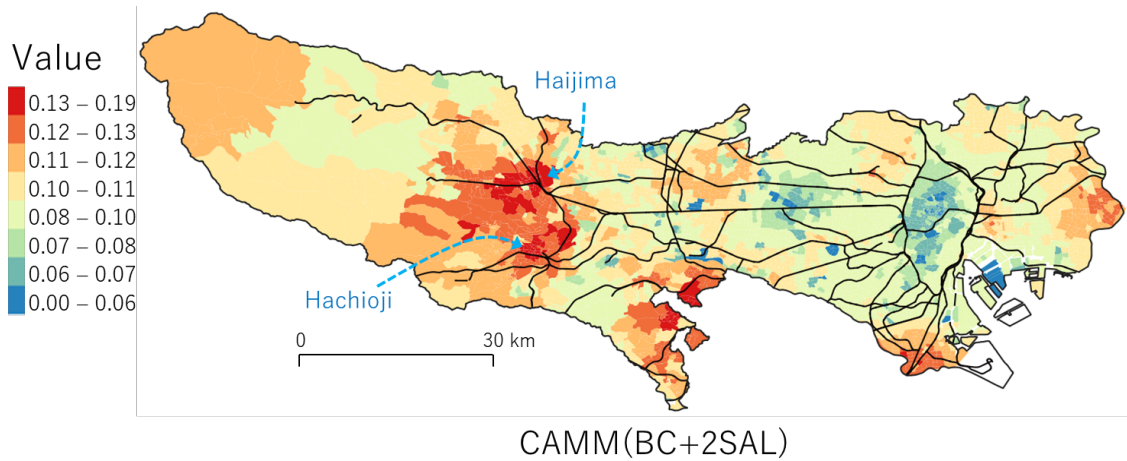

CAMM(BC+2SAL)

Figure 12: Estimated SNVC on RepOther. The black lines are railways not including subways.

Finally, Table 3 summarizes the estimated median marginal effects. For example, the marginal effect from Repeat of $5.714\times10^{-2}$ means that shoplifting per retail (i.e., explained variables value) is expected to increase $5.714\times10^{-2}$ if Repeat increases by 1. The marginal effects estimates, which quantify the influence on the raw explained variables $y_i$, might be more interpretable than the naïve coefficient estimates measuring the impact on $\varphi_{\boldsymbol{\omega}}(y_i)$. Table 3 demonstrates that the selection of the transformation function significantly changes the marginal effects estimates. Overall, CAMM tends to have similar or greater marginal effects from significant variables (Repeat, RepOther, Popden, Retail) and similar or smaller marginal effects from insignificant variables (Fpopden, UnEmp, Univ). Based on the better BIC value, CAMM must be used to accurately measure the impact from each explanatory variable.

Table 3: Medians of the estimated marginal effects
($\partial y_i/\partial x_{i,k}$; impact of a unit increase of $x_{i,k}$ on the number of shoplifting occurrences per retail).

|          | AMM(log)              | CAMM(BC+2SAL)         |
|----------|-----------------------|-----------------------|
| Repeat   | $5.280\times10^{-2}$  | $5.714\times10^{-2}$  |
| RepOther | $6.876\times10^{-3}$  | $8.032\times10^{-3}$  |
| Popden   | $-7.678\times10^{-5}$ | $-7.091\times10^{-5}$ |
| Retail   | $6.708\times10^{-7}$  | $1.143\times10^{-6}$  |
| Fpopden  | $-3.682\times10^{-3}$ | $-5.741\times10^{-3}$ |
| UnEmp    | $2.776\times10^{-3}$  | $3.294\times10^{-4}$  |
| Univ     | $2.330\times10^{-3}$  | $3.685\times10^{-4}$  |

6.4. Application to crime prediction

Finally, the models estimated using the data between 2015 and 2018 by quarter are used to predict the number of shoplifting per retail in the first quarter in 2019. Note that Murakami et al. (2020) demonstrated the higher prediction accuracy of AMM compared to the kernel density estimation, which is a popular crime mapping method. The objective here is to examine if our proposed transformation further improves the prediction accuracy.

Figure 13 mapped the observed and predicted number of shoplifting occurrences per retail. This figure shows that both seemingly reproduce accurate distribution of the crimes. Figure 14 is a plot with the x- and y-axis representing the observed and predicted values, respectively. AMM(log) tends to underestimate in the large value region. On the other hand, CAMM(BC+2SAL) accurately predicts both small and large observations. The difference might be caused by the fact that CAMM accurately models the upper tail of the data distribution using the SAL transformation while AMM failed to achieve it (see Figure 10). In fact, the root mean squared prediction error (RMSPE) of CAMM is 1.461 that is smaller than AMM (RMSPE: 1.604).

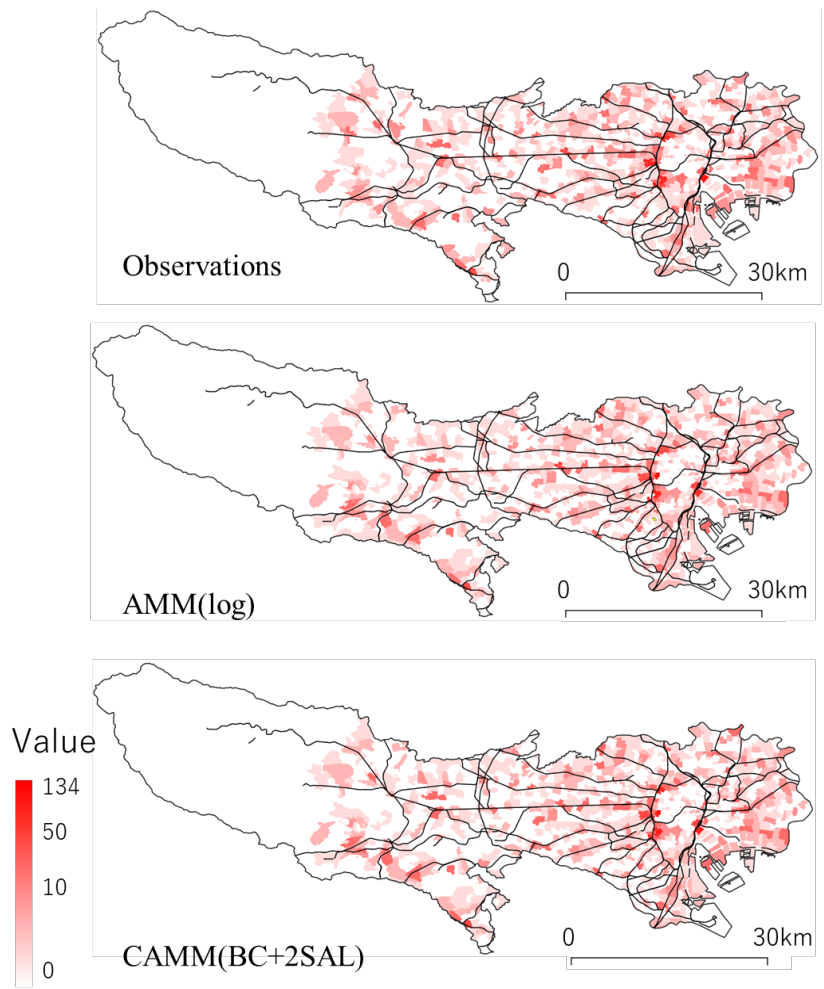

Figure 13: Observed and predicted number of shoplifting occurrences per retail in the first quarter, 2019.

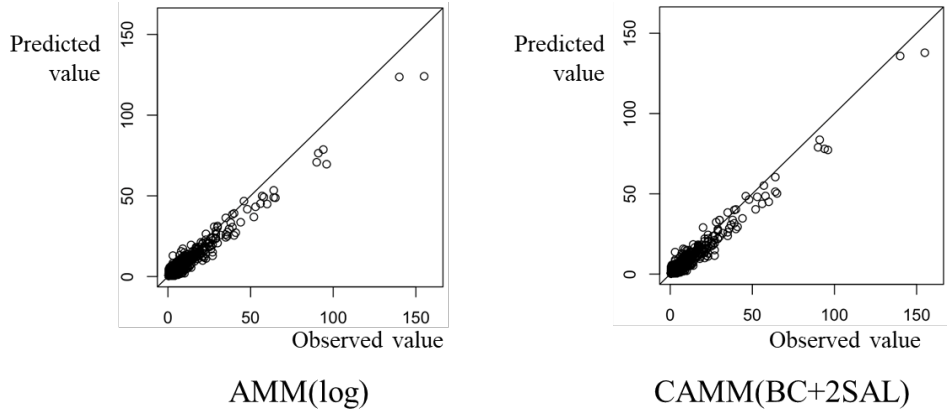

AMM(log)          CAMM(BC+2SAL)

Figure 14: Observed and predicted number of shoplifting occurrences per retail in the first quarter, 2019.

## 7. Concluding remarks

This study proposed the compositionally-warped additive mixed modeling (CAMM), which is a general framework for fast non-Gaussian regression modeling. Unlike other non-Gaussian additive models, CAMM does not require any explicit assumptions on data distribution. CAMM will be useful to estimate spatial, temporal, and other effects while avoiding misspecification relating to data distribution. The Monte Carlo experiments verify the accuracy and computational efficiency of our approach, and the empirical part confirms its usefulness in practice.

The developed approach is potentially applicable to a wide variety of spatial and spatiotemporal modeling including hedonic price modeling (e.g., Brasington and Hite, 2005; Lu et al., 2014b) and other econometric modeling (e.g., Autant-Bernard and LeSage, 2011), environmental modeling (e.g., Gillbert and Chakraborty, 2011), ecological modeling (e.g., Bini et al., 2009), and modeling diseases such as COVID-19 (e.g., Sannigrahi et al., 2020). Many more applications are needed to clarify how and when the developed approach is useful.

Despite this, as with CWGP, the developed CAMM limits its application only to continuous explained variables. For modeling discrete variables like counts, they must be divided by area or population to convert to density. Although such division in itself can be included as another transformation under the CWGP/CAMM framework, it is more straightforward to model discrete variable directly. Lázaro-Gredilla (2012) develops a Bayesian WGP accommodating both continuous and discrete variables. Bayesian extension will be required to extend CAMM for discrete data modeling. Such an extension will also be valuable to consider uncertainty of the transformation.

Extension of the transformation function is another potential future work. Although the transformations are done for each sample independently, dependence is potentially introduced across samples. For example, the number of transformations and parameter values for the transformations can be spatially varying to capture spatial non-stationarity. Given that the iterative SAL transformation can be viewed as a neural network with single node in each layer (see Rois and Tober, 2019), deep learning algorithm, especially the algorithms of deep GP (Damianou and Lawrence, 2013), might be useful to model highly complex spatial and spatiotemporal processes (see Zammit-Mangion et al., 2019).

CAMM was implemented in the R package spmoran (https://cran.r-project.org/web/packages/spmoran/index.html). For details on how to implement CAMM, see https://github.com/dmuraka/spmoran.

# Acknowledgements

This work was supported by JSPS KAKENHI Grant Numbers 17H02046, 18H03628, 20K13261. Also, these research results were obtained from the research commissioned by the National Institute of Information and Communications Technology (NICT), Japan.

# Appendix 1. The restricted log-likelihood maximization approach

In the estimation steps (3) – (5), $\log L_R(\mathbf{\Theta}, \boldsymbol{\omega}, \sigma^2) = \log L_R^{AMM}(\mathbf{\Theta}, \boldsymbol{\omega}, \sigma^2) + \sum_{i=1}^{N} \log \frac{\partial \varphi_\omega(y_i)}{\partial y_i}$ is maximized given $\widetilde{\mathbf{M}} \in \{\mathbf{M}_{XX}, \mathbf{M}_{EX}, \mathbf{M}_{EE}, \mathbf{m}_{Xy}, \mathbf{m}_{Ey}, m_{yy}\}$. The computational complexity of $\sum_{i=1}^{N} \log \frac{\partial \varphi_\omega(y_i)}{\partial y_i}$ is linear with respect to the samples size (see Rois and Tober, 2019). The computational cost is trivial.

Regarding $\log L_R^{AMM}(\mathbf{\Theta}, \boldsymbol{\omega}, \sigma^2)$, which equals the log-restricted likelihood of the AMM. Murakami and Griffith (2019; 2020a) showed that it can be written as

$$\log L_R^{AMM}(\mathbf{\Theta}, \boldsymbol{\omega}, \sigma^2) = -\frac{1}{2}\log|\mathbf{R}(\mathbf{\Theta})| - \frac{N-K}{2}\left(1 + \log\left(\frac{2\pi d(\mathbf{\Theta})}{N-K}\right)\right), \quad (A1)$$

where

$$\boldsymbol{d}(\mathbf{\Theta}) = m_{y,y} - 2[\hat{\mathbf{b}}', \hat{\mathbf{u}}']\begin{bmatrix}\mathbf{m}_{Xy}\\\widetilde{\mathbf{V}}(\mathbf{\Theta})\mathbf{m}_{Ey}\end{bmatrix} + [\hat{\mathbf{b}}', \hat{\mathbf{u}}']\begin{bmatrix}\mathbf{M}_{XX} & \mathbf{M}_{EX}\widetilde{\mathbf{V}}(\mathbf{\Theta})\\\widetilde{\mathbf{V}}(\mathbf{\Theta})\mathbf{M}_{EX} & \widetilde{\mathbf{V}}(\mathbf{\Theta})\mathbf{M}_{EE}\widetilde{\mathbf{V}}(\mathbf{\Theta})\end{bmatrix}\begin{bmatrix}\hat{\mathbf{b}}\\\hat{\mathbf{u}}\end{bmatrix} + \|\hat{\mathbf{u}}\|^2, \quad (A2)$$

$$\begin{bmatrix}\hat{\mathbf{b}}\\\hat{\mathbf{u}}\end{bmatrix} = \begin{bmatrix}\mathbf{M}_{XX} & \mathbf{M}_{EX}\widetilde{\mathbf{V}}(\mathbf{\Theta})\\\widetilde{\mathbf{V}}(\mathbf{\Theta})\mathbf{M}_{EX} & \widetilde{\mathbf{V}}(\mathbf{\Theta})\mathbf{M}_{EE}\widetilde{\mathbf{V}}(\mathbf{\Theta}) + \mathbf{I}\end{bmatrix}^{-1}\begin{bmatrix}\mathbf{m}_{Xy}\\\widetilde{\mathbf{V}}(\mathbf{\Theta})\mathbf{m}_{Ey}\end{bmatrix}, \quad (A3)$$

The computational complexity of evaluating $\log L_R^{AMM}(\mathbf{\Theta}, \boldsymbol{\omega}, \sigma^2)$ is independent of *N* because Eqs. (A1) – (A3) do not include any matrix or vector whose size growth with respect to *N*. The computation cost for Eqs. (A1) is $O(\tilde{L}^3)$, which is needed for the inversion in Eq. (A3); the computation cost is quite small. For further detail about the REML for AMM, see Murakami and Griffith (2019).

After all, the REML maximizing $\log L_R(\mathbf{\Theta}, \boldsymbol{\omega}, \sigma^2)$ will be fast even for large samples.

# Appendix 2. Transformation functions used in this study

The SAL transformation in each $d \in \{1, \ldots, D\}$, it can be replaced with other transformation functions. The transformation functions considered in this study are summarized in Table B1.

Table B1: Transformation functions considered in this study

| | $\varphi_{\boldsymbol{\omega}_d}(y_i)$ | $\varphi_{\boldsymbol{\omega}_d}^{-1}(y_i)$ | $\dfrac{\partial \varphi_{\boldsymbol{\omega}_d}(y_i)}{\partial y_i}$ |
|---|---|---|---|
| SAL trans. | $\omega_{d,1} + \omega_{d,2}\sinh(\omega_{d,3}\arcsinh(y_i) - \omega_{d,4})$ | $\sinh\left(\dfrac{1}{\omega_{d,3}}\left(\operatorname{asinh}\left(\dfrac{y_i - \omega_{d,1}}{\omega_{d,2}}\right) + \omega_{d,4}\right)\right)$ | $\dfrac{\omega_{d,2}\omega_{d,3}\cosh(\omega_{d,3}\operatorname{asinh}(y_i) - \omega_{d,4})}{\sqrt{1 + y_i^2}}$ |
| Box-Cox trans. | $\begin{cases} \dfrac{y_i^{\omega_d} - 1}{\omega_d} & if\ \omega_d = 0 \\ \log(y_i) & Otherwise \end{cases}$ | $\begin{cases} (\omega_d y_i + 1)^{\frac{1}{\omega_d}} & if\ \omega_d = 0 \\ \exp(y_i) & Otherwise \end{cases}$ | $y_i^{\omega_d - 1}$ |
| Log trans. | $\log(y_i)$ | $\exp(y_i)$ | $y_i^{-1}$ |
| Standardization | $\dfrac{y_i - \omega_{d,1}}{\omega_{d,2}}$ | $\omega_{d,2} y_i + \omega_{d,1}$ | $\omega_{d,2}^{-1}$ |
| Add value | $y_i + \omega_{d,1}$ | $y_i - \omega_{d,1}$ | 1 |

# Appendix. 3: Implementation using the spmoran package

This appendix explains how to implement the CAMM using an R package "spmoran" (version 0.2.1 onward). Here is the code used in the analysis:

```
meig    <- meigen( coords=coords, s_id=s_id )
mod     <- resf_vc( meig=meig, y=y, x=x, xgroup = xgroup, x_nvc=TRUE,
                    tr_num=2, tr_nonneg = TRUE )
me      <- coef_marginal( mod )
```

where meigen is the function extracting Moran eigenvectors, which are used to specify the spatially varying intercept and SVCs. Spatial coordinates (coords) are used for the extraction (see Drey et al., 2006; Murakami and Griffith, 2015). If location ID (s_id) is specified, the Moran eigenvectors are extracted by the ID, which is given by minor municipal district in our case. The meigen function, implementing an eigen-decomposition that is slow for large samples, can be replaced with the meigen_f function, which approximates the Moran eigenvectors.

The resf_vc function by default estimates a linear SVC model, which is popular in applied spatial statistics (e.g., Brunsdon et al., 1998). This function is available to estimate the CAMM with varying coefficients. The inputs are as follows:

- meig   : The output from the meigen or meigen_f function
- y      : Vector of explained variables
- x      : Matrix of explanatory variables
- xgroup : One or more group IDs for defining random intercept (ID by quartile in our case)
- x_nvc  : If FALSE, coefficients are defined by SVCs. If TRUE, they are defined by SNVC. The latter specification is much more stable than naïve SVC models (see Murakami and Griffith., 2020b).
- tr_num : The number of SAL transformations
- tr_nonneg : If TRUE, the Box-Cox transformation is applied first, and the SAL transformations after that (i.e., B in Figure 2).

By default, this function selects coefficients type through a BIC minimization (among {Const, SVC} if x_nvc=FALSE while {Const, SVC, NVC, SNVC} if x_nvc=TRUE). When assuming constant coefficients just like standard regression analysis, the resf function is available. Lastly, the coef_marginal function returns summary statistics of the estimated marginal effects. See Murakami (2020) for further detail.